\documentclass[twocolumn,twoside]{revtex4}
\usepackage{graphicx}
\usepackage{fancyhdr}
\begin{document}

\title{Electroweak Physics with Polarized Electron Beams in a SuperKEKB Upgrade}

%

\author{J.~Michael~Roney\\}
\affiliation{University of Victoria, BC, Canada, V8W 2Y2}

\begin{abstract}
Consideration is being given to upgrading the SuperKEKB e$^+$e$^-$
 collider with polarized electron beams, which would open a new program of precision electroweak physics at a centre-of-mass energy of 10.58~GeV, the mass of the $\Upsilon(4S)$.  
These measurements include $\sin^2\theta_W$ obtained via left-right asymmetry measurements of e$^+$e$^-$ transitions to pairs of electrons, muons, taus, charm and b-quarks.
 The precision obtainable at SuperKEKB will match that of the LEP/SLC world average and will thereby probe the neutral current couplings with unprecedented precision
 at a new energy scale sensitive to the running of the couplings.  At SuperKEKB the measurements of the individual neutral current vector coupling constants
 to b-quarks and c-quarks and muons in particular will be substantially more precise than current world averages and the current 3$\sigma$
 discrepancy between the SLC A$_{LR}$ measurements and LEP A$_{FB}^b$ measurements of $\sin^2\theta_W^{eff}$
will be addressed. This paper will include a discussion of the necessary upgrades to SuperKEKB.
 This program opens an exciting new window in searches for physics beyond the Standard Model.

\end{abstract}

\maketitle

\thispagestyle{fancy}


With its high design luminosity of $8\times 10^{35}$~cm$^{-2}$~s$^{-1}$,
the SuperKEKB e$^+$e$^-$ collider operating at a centre-of-mass energy of 10.58~GeV can access new windows for discovery with the Belle~II 
experiment if it is upgraded to have a longitudinally polarized electron beam.
 The target integrated luminosity for SuperKEKB/Belle~II is 50~ab$^{-1}$\cite{BelleIITDR}
 and currently Belle~II is projected to collect that amount of data, which will not have beam polarization without an upgrade to SuperKEKB, in 2027.
 Upgrading SuperKEKB to have electron beams with left and right longitudinal polarization of approximately 70\% at the Belle~II interaction point creates 
 a unique and versatile facility for probing new physics with precision electroweak measurements that no other experiments, current or planned, 
 can achieve.

 In particular, a data sample of 20~ab$^{-1}$ with a polarized electron beam enables Belle~II to measure the weak neutral current vector coupling 
 constants of the $b$-quark, $c$-quark and muon at significantly higher precision than any previous experiment. 
 With 40~ab$^{-1}$ of polarized beam data, the precision of
 the vector couplings to the tau and electron can be measured at a level comparable to current world averages, which are dominated by
 LEP and SLD measurements at the Z$^0$-pole.

 Within the framework of the Standard Model (SM) these measurements of $g_V^f$, the neutral current vector coupling for fermion $f$,
 can be used to determine the weak mixing angle, $\theta_W$, 
through the relation: $g_V^f = T_3^f - 2Q_f \sin^2\theta_W$, where $T_3^f$ is the $3^{rd}$ component of weak isospin of fermion $f$,
 $Q_f$ is its electric charge in units of electron charge and
the notational conventions of Reference~\cite{LEPSLDReport2006} are used.

As described in Reference~\cite{SuperBTDR},
with polarized electron beams an e$^+$e$^-$ collider at 10.58~GeV can
 determine $g_V^f$  by measuring the left-right asymmetry, $A_{LR}^f$, for each identified final-state fermion-pair in the process
$e^+e^- \rightarrow f \overline{f}$:
\begin{equation}
 A_{LR}^f = \frac{ \sigma_L - \sigma_R}{\sigma_L + \sigma_R} = \frac{s G_F }{\sqrt{2} \pi \alpha Q_f} g_A^e g_V^f \langle Pol \rangle
\end{equation}
where $g_A^e = T_3^e = -\frac{1}{2}$ is the neutral current axial coupling of the electron, $G_F$ is the Fermi coupling constant,
$s$ is the square of the centre-of-mass energy, and 
\begin{equation}
\langle Pol \rangle =\frac{1}{2} \left[ \left(\frac{N_{eR}-N_{eL}}{N_{eR}+N_{eL}}\right)_{\mathbf R} -
 \left(\frac{N_{eR}-N_{eL}}{N_{eR}+N_{eL}}\right)_{\mathbf L} \right]
\end{equation}
is the average electron beam polarization, where  $N_{eR}$ is the number of right-handed electrons and $N_{eL}$
 the number of left-handed electrons in the event samples where the electron beam bunch is left polarized or right polarized,
 as indicated by the `${\mathbf R}$' and `${\mathbf L}$' subscripts. 
 These asymmetries arise from $\gamma-Z$ interference~\cite{Bernabeu} and although the
 SM asymmetries are small ($-6\times 10^{-4}$ for the leptons, $-5\times 10^{-3}$ for charm and
 $-2\%$ for the $b$-quarks), unprecedented precisions can be achieved 
 because of the combination of both the high luminosity of SuperKEKB and a $70\%$ beam polarization measured with a precision of better than $\pm 0.5\%$.
Note that higher order corrections are ignored here for simplicity, although studies at higher orders have recently begun~\cite{Aleskejevs}.

 The upgrade to SuperKEKB involves three hardware projects:
\begin{itemize}
\item A low-emittance polarized electron source in which 
  the electron beams would be produced via a polarized laser illuminating a ``strained lattice'' GaAs photocathode
 as was done for SLD~\cite{LEPSLDReport2006}. The source would produce longitudinally polarized electron bunches whose spin would be rotated to be transversely polarized before it
 enters the SuperKEKB electron storage ring;
\item A pair of spin-rotators, one positioned before and the other after the interaction region, to rotate the spin to longitudinal
 prior to collisions and back to transverse following collisions. One configuration under consideration for the spin-rotator is a
  combined function magnet that replaces an existing dipole in the SuperKEKB electron beam lattice with a  magnet that is both a dipole and solenoid~\cite{UliWienands}.
 The challenge is to design the rotators to minimize couplings between vertical and horizontal planes and to address higher order and chromatic effects 
 in the design to ensure the luminosity is not degraded.
\item A Compton polarimeter that measures the beam polarization before the beam enters the interaction region.
\end{itemize}

 The high precisions are possible at such an upgraded SuperKEKB
 because with 20~ab$^{-1}$ of data Belle~II can identify between $10^9$ and $10^{10}$ final-state pairs of b-quarks, c-quarks, taus, muons and electrons
 with high purity and reasonable signal efficiency,
 and because all detector-related systematic errors can be made to cancel by flipping
 the laser polarization from ${\mathbf R}$ to ${\mathbf L}$ in a random, but known, pattern as collisions occur.
 $\langle Pol \rangle$ would be measured in two ways. The first method uses a Compton polarimeter, which can be expected to have an absolute
 uncertainty at the Belle~II interaction point of $<1\%$ and provides a `bunch-by-bunch' measurement of $\left(\frac{N_{eR}-N_{eL}}{N_{eR}+N_{eL}}\right)_{\mathbf R}$
 and $\left(\frac{N_{eR}-N_{eL}}{N_{eR}+N_{eL}}\right)_{\mathbf L}$. The uncertainty will be dominated by the need to predict the polarizatoin loss
 from the Compton polarimeter to the interaction point. The second method 
 measures the polar angle dependence of the polarization of $\tau$-leptons produced in $e^+e^- \rightarrow \tau^+ \tau^-$ events using the kinematic distributions of the
 decay products of the $\tau$ separately for the ${\mathbf R}$ and ${\mathbf L}$ data samples.
 The forward-backward asymmetry of the tau-pair polarization is linearly dependent on $\langle Pol \rangle$ and therefore
 can be used to determine $\langle Pol \rangle$ to $0.5\%$ at the Belle~II interaction point in a manner entirely independent of the Compton polarimeter. 
 The $\tau$ polarization forward-backward asymmetry method avoids the uncertainties associated with tracking the polarization losses to the interaction point and also
 automatically accounts for any residual positron polarization that might be present. In addition, it automatically provides a luminosity-weighted beam
 polarization measurement.

 Table~\ref{Table-measurements} provides the sensitivities to electroweak parameters expected with polarized electron beams in an upgraded SuperKEKB from
$e^+e^- \rightarrow b \bar{b}$, $e^+e^- \rightarrow c \bar{c}$, $e^+e^- \rightarrow \tau^+ \tau^-$, $e^+e^- \rightarrow \mu^+ \mu^-$, 
and $e^+e^- \rightarrow e^+ e^-$ events selected by Belle~II. From this information the precision on the b-quark, c-quark and muon neutral current vector couplings 
will improve by a factor of four, seven and five, respectively, over the current world average values\cite{LEPSLDReport2006} with 20~ab$^{-1}$ of polarized data.

This is of particular importance for $g^b_V$, where the measurement of -0.3220$\pm$0.0077 is  2.8$\sigma$ higher than the SM value of -0.3437~\cite{LEPSLDReport2006}.
That discrepancy is a manifestation of the 3$\sigma$ difference between the SLC $A_{LR}$ measurements and LEP $A^b_{FB}$ measurements
 of $\sin^2\theta_W^{eff}$~\cite{LEPSLDReport2006}. 
A measurement of $g^b_V$ at an upgraded SuperKEKB that is four times more precise and which avoids the hadronization uncertainties that are a significant component
 of the uncertainties of the measurement of $A^b_{FB}$ at LEP, will be able to definitively resolve whether or not this is
a statistical fluctuation or a first hint of a genuine breakdown of the SM. 

\begin{figure*}[h]
\centering
\includegraphics[width=135mm]{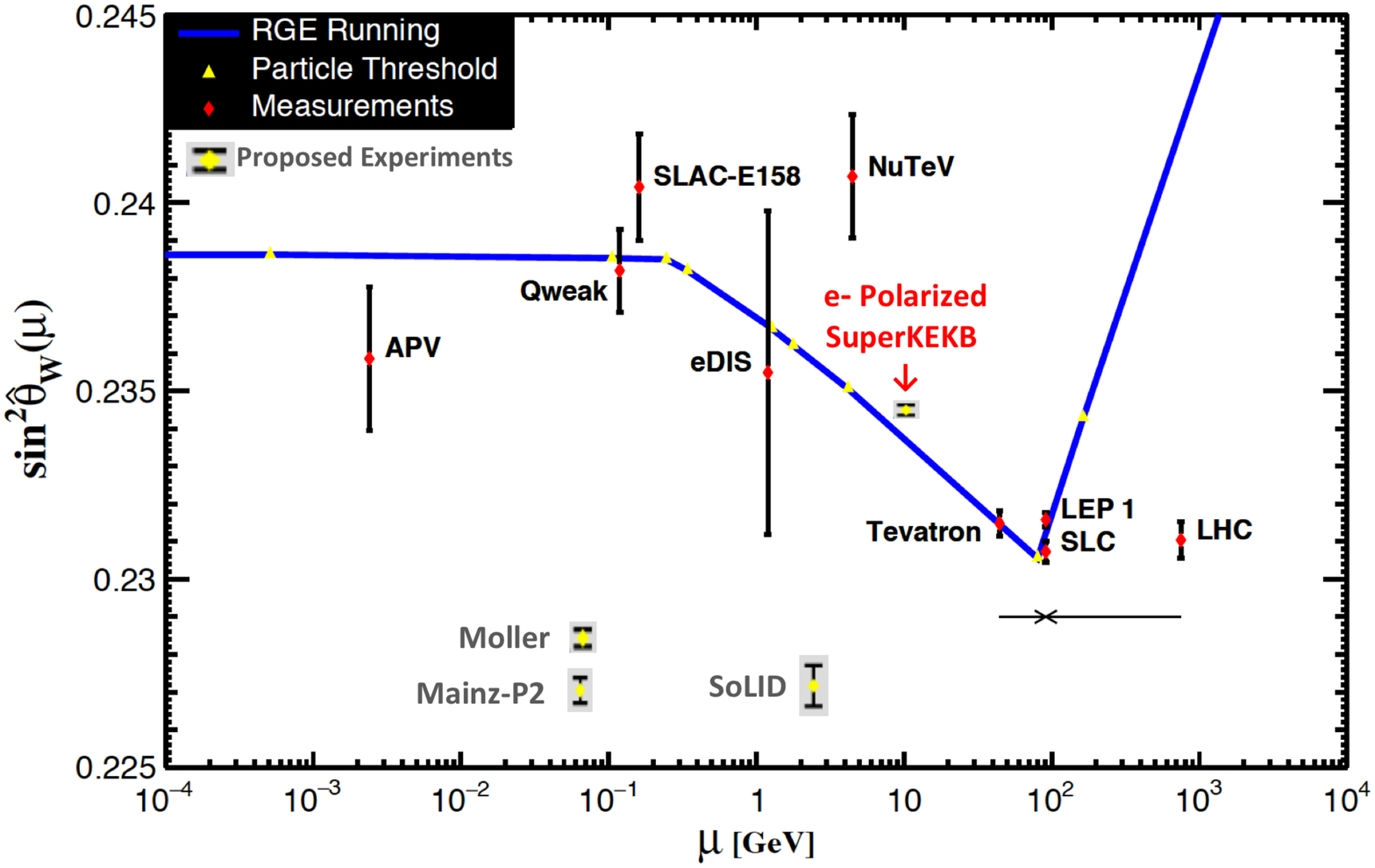}
\caption{Scale dependence of $\sin^2\theta_W$ in the $\overline{MS}$ renormalization scheme~\cite{ErlerPDG2018}.
Present measurements are shown with red diamonds. The uncertainties and energy scale of 
 future experimental facilities are added to the figure  in \cite{ErlerPDG2018} as yellow diamonds.
The SuperKEKB point assumes 40~ab$^{-1}$ of data.}
 \label{fig:sin2thetaW}
\end{figure*}

\begin{table*}[h]
\begin{center}
\caption{For each fermion pair that can be cleanly identified in Belle~II for the given efficiency in column~1: column~2 gives
the SM value of the left-right asymmetry; column~3 gives the expected 
relative error on an upgraded SuperKEKB/Belle~II measurement of $A_{LR}$ calculated as the quadratic sum of
 the statistical error (based on 20~ab$^{-1}$, the assumed selection efficiencies given in column~1 and
 a beam polarization at Belle~II of 70\%) and a systematic error of $\pm0.5\%$ on the 70\% beam polarization;
column~4 gives the current world average value of its neutral current vector coupling;
column~5 gives the projected error on $g_V^f$ for an upgraded SuperKEKB/Belle~II measurement using 20~ab$^{-1}$ of data; 
column~6 gives the projected error on $g_V^f$ using 40~ab$^{-1}$ of data; and 
column~7 gives the projected SuperKEKB/Belle~II error on $\sin^2\theta_W^{eff}$ with 40~ab$^{-1}$ of polarized beam data.
}
\begin{tabular}{|l|c|c|r|c|c|c|}
\hline
Final State & $A_{LR}^{SM}$ &  Relative           &$g^f_V$ W.A.\cite{LEPSLDReport2006} & $\sigma(g^f_V)$  & $\sigma(g^f_V)$  & $\sigma (\sin^2\theta_W^{eff})$ \\
Fermion     &               &  $A_{LR}$ Error (\%)&                        & for 20~ab$^{-1}$ & for 40~ab$^{-1}$ &  for 40~ab$^{-1}$             \\ 
\hline 
 b-quark     &  -0.020      &  0.5                &  -0.3220               &   0.002          &   0.002          & 0.003 \\
 (eff.=0.3)  &              &                     & $\pm 0.0077$           &improves x4        &                  &       \\
 \hline
 c-quark     &  -0.005      &  0.5                &  +0.1873               &   0.001          &   0.001          & 0.0007 \\
 (eff.=0.3)  &              &                     &  $\pm 0.0070$          &  improves x7      &                  &        \\
 \hline
 tau         & -0.0006      &  2.3                &  -0.0366               &   0.0008         &   0.0006         &  0.0003 \\
 (eff.=0.25) &              &                     &  $\pm 0.0010$          &                  &                  &       \\ 
 \hline
 muon        & -0.0006      &  1.5                &  -0.03667              &   0.0005         &   0.0004         & 0.0002 \\
 (eff.=0.5)  &              &                     &  $\pm 0.0023$          &   improves x5     &                  &       \\
 \hline
 electron    & -0.0006      &  1.2                &  -0.3816               &   0.0004         &   0.0003         & 0.0002 \\
(1~nb acceptance)&          &                     &  $\pm 0.00047$         &                  &                  &       \\
 \hline
\end{tabular}
\label{Table-measurements}
\end{center}
\end{table*}

Table~\ref{Table-measurements} also indicates the uncertainties on $\sin^2\theta_W^{eff}$ that can be achieved with
 40~ab$^{-1}$ of polarized beam data - the combined uncertainty at Belle~II would be comparable to the Z$^0$ world average measured uncertainty
of $\pm 0.00016$ from LEP and SLD\cite{LEPSLDReport2006} but made at a significanlty lower energy scale.
Figure~\ref{fig:sin2thetaW} shows the determinations of $\sin^2\theta_W$ in the $\overline{MS}$ renormalization scheme
 as a function of energy scale~\cite{ErlerPDG2018} at present and future experimental
facilities including SuperKEKB upgraded with a polarized electron beam delivering 40~ab$^{-1}$ of data to Belle~II.

This electroweak program with polarized electron beams in SuperKEKB
 would provide the highest precision tests of neutral current vector coupling universality.
In addition, right-handed $b$-quark couplings to the $Z$ can  be experimentally
probed with high precision at Belle~II with polarized beams.  Also, measurements with the
projected precision will enable Belle~II to probe parity violation
induced by the exchange of heavy particles such as a
hypothetical TeV-scale $Z^\prime$ boson(s). If such bosons only couple
to leptons they will not be produced at the LHC.  Moreover, the
SuperKEKB machine will have a unique possibility to probe parity
violation in the lepton sector mediated by light and very weakly
coupled particles often referred to as ``Dark Forces''.
Such forces have been entertained as a possible connecting link between normal and dark
matter \cite{Pospelov:2008jd,ArkaniHamed:2008qn}. 
SuperKEKB with polarization would be uniquely sensitivity to ``Dark Sector'' parity violating light neutral gauge
bosons – especially when $Z_{dark}$ is
 off-shell and with a mass between roughly 10 and 35~GeV \cite{Davoudiasl} or even up to the Z$^0$ pole,
 or couples more to the 3rd generation.

The enhancement of
parity violation in the muon sector has been an automatic consequence
of some models \cite{Batell:2011qq} that aim at explaining the
unexpected result for the recent Lamb shift measurement in muonic
hydrogen \cite{Pohl:2010zza}. The left-right asymmetry of the $e^-e^+
\to \mu^-\mu^+$ in such models is expected to be enhanced at
low-to-intermediate energies, and SuperKEKB with polarized beams may provide a
conclusive test of such models, as well as impose new constraints on a
parity-violating dark sector.

An electron beam polarization upgrade at SuperKEKB  opens 
an exciting and unique discovery window with precision electroweak physics. 
 A growing international team  has begun to study the feasibility of such an upgrade with the
goal to begin taking Belle~II data with polarized SuperKEKB electron beams in the mid-2020’s.

\begin{acknowledgments}
The collaboration with Uli Wienands and Demin Zhou have been essential for developing the accelerator concepts needed for a potential
beam polarization upgrade to SuperKEKB. Ongoing work by Caleb Miller on the measurement of tau polarization at 10.58~GeV is also
gratefully acknowledged.
\end{acknowledgments}

\bigskip 

\end{document}